\documentclass[twocolumn]{aastex631} 

\usepackage{amsmath}
\usepackage{url}
\usepackage{lipsum} 
\usepackage{graphicx}
\usepackage{multirow}

\begin{document}

\title{The Delta Resonance in the Neutrino Sky}

\author[0000-0002-8735-8579]{Arifa Khatee Zathul}
\affiliation{Department of Physics, Wisconsin IceCube Particle Astrophysics Center, University of Wisconsin–Madison, Madison, WI 53706, USA}

\author[0000-0002-5387-8138]{Ke Fang}
\affiliation{Department of Physics, Wisconsin IceCube Particle Astrophysics Center, University of Wisconsin–Madison, Madison, WI 53706, USA}

\author[0000-0001-6224-2417]{Francis Halzen}
\affiliation{Department of Physics, Wisconsin IceCube Particle Astrophysics Center, University of Wisconsin–Madison, Madison, WI 53706, USA}

\author[0009-0004-2456-1221]{Dan Hooper}
\affiliation{Department of Physics, Wisconsin IceCube Particle Astrophysics Center, University of Wisconsin–Madison, Madison, WI 53706, USA}
 

\begin{abstract}

Recent measurements of the diffuse cosmic neutrino flux by IceCube show evidence for a spectral break at an energy near $E_\nu \sim 30$ TeV. In this letter, we suggest that this feature may be due to the $\Delta$-baryon resonance in $p\gamma$ interactions. We show that the measured spectrum, including the observed break, can be naturally accommodated by a flux of protons accelerated with a spectrum $dN_p /dE_p \propto E_p^{-3.1}$ interacting with X-rays of typical energy $E_{\gamma} \sim 0.3\,{\rm keV}$. We also point out that the presence of this spectral break  reduces the contribution of neutrino sources to the isotropic gamma-ray background, alleviating the longstanding tension between these measurements. In the $\Delta$-resonance scenario, the gamma rays accompanying neutrino production cascade down to MeV-GeV energies and contribute at the $\sim 10\%$ level to the isotropic gamma-ray background at $\sim 3$~GeV. If our proposal is realized, it may imply that we have identified the dominant sources that produce the extragalactic cosmic rays.

\end{abstract}


\section{Introduction}

The IceCube Neutrino Observatory has identified the first astrophysical sources of high-energy neutrinos, including the nearby active galaxy NGC 1068~\citep{icecube2022evidence}, the blazar TXS 0506+056~\citep{IceCube:2018cha,luszczak2023txs}, and the Galactic Plane~\citep{IceCube:2023ame}. In addition to these individual sources, IceCube's measurement of the diffuse high-energy neutrino flux~\citep{PhysRevLett.111.021103, icecube2013evidence, 
PhysRevLett.113.101101} provides key constraints on the source populations responsible for producing these neutrinos.

Since the discovery of the diffuse flux in 2013, the IceCube Collaboration has refined its measurements of this signal, using larger datasets and improved event selection and reconstruction~\citep{aartsen2019measurements, aartsen2020characteristics, abbasi2021icecube, abbasi2022improved, abbasi2026time}. These studies have inferred a spectrum that is broadly consistent with a power law with spectral index $\gamma \sim 2.3$–$2.9$. More recent analyses, however, indicate that the diffuse spectrum softens significantly in the TeV-PeV energy range and exhibits a feature at around $\sim 30$ TeV~\citep{abbasi2025improved}. In this letter, 
we investigate the origin of this spectral break and discuss its implications for high-energy, multimessenger astrophysics. 

We propose that the spectral feature observed in the diffuse neutrino spectrum may arise from proton interactions with radiation proceeding through the $\Delta$-baryon resonance, $p + \gamma \rightarrow \Delta \rightarrow n + \pi^+$. This resonance is expected to play an important role in environments with high densities of X-ray radiation, including those surrounding active galactic nuclei such as NGC 1068.

Pions produced through the $\Delta$ resonance carry $\sim 20\%$ of the energy of the initial proton on average. Combined with the fact that each neutrino carries, on average, about one quarter of the pion energy in the decay chain $\pi^+ \rightarrow \mu^+ + \nu_\mu$ followed by $\mu^+ \rightarrow e^+ + \nu_e +\bar\nu_\mu$, this process yields neutrinos with $E_\nu \sim 0.05 \, E_p$. 

For interactions on the $\Delta$ resonance, the energy in the center-of-momentum frame satisfies the following condition:
\begin{align}
m_{\Delta}^2 = E^2_{\rm CM} = 2 E_p E_{\gamma} +m^2_p - 2 E_{\gamma} \sqrt{E^2_p-m^2_p} \, \cos \theta,
\end{align}
where $\theta$ is the angle between the momenta of the proton and the target photon. Averaging over this angle, we obtain
\begin{align}
&m_{\Delta}^2 -m^2_p \approx  2 E_p E_{\gamma} \langle 1 - \cos \theta \rangle,
\end{align}
and thus 
\begin{align}
E_p E_{\gamma} \approx \frac{m_{\Delta}^2 -m^2_p}{2} \approx 0.3 \, {\rm GeV}^2.
\end{align}
Producing a spectral feature at $E_{\nu} \sim 30 \, {\rm TeV}$ therefore requires protons with energies $E_p \sim 0.6 \, {\rm PeV}$ and target photons in the $\sim 0.1$-$1 \, {\rm keV}$ range (neglecting energy losses due to cosmological redshift).

Neutrino production via pion decay is necessarily accompanied by a comparable flux of gamma rays from the reaction $p + \gamma \rightarrow p + \pi^0$. At energies above $\sim 200 \, {\rm GeV}$, gamma rays are attenuated over cosmological distance scales through electron-positron pair production on the extragalactic background light (EBL), followed by inverse Compton scattering of the resulting electrons and positrons. This process generates an electromagnetic cascade that ultimately results in a background of photons at GeV-TeV energies. 

It has long been recognized that if the diffuse neutrino spectrum extends to energies below $\sim 1-10 \, {\rm TeV}$ without a significant spectral break, the associated gamma-ray emission would exceed, or otherwise be in significant tension with, measurements of the isotropic gamma-ray background (IGRB)~\citep{Murase:2015xka, Hooper:2016jls, fang2022tev}, unless the sources of these neutrinos are largely opaque to gamma rays. As we will show here, the $\Delta$ resonance provides a natural mechanism for generating such a spectral break, thereby relieving this tension with the measured IGRB.

In the letter, we show that the diffuse neutrino flux measured by IceCube, including the spectral break at $\sim30$ TeV, can be naturally explained by protons accelerated with a spectrum of $dN_p/dE_p \propto E_p^{-3.1}$ interacting with X-rays of typical energy $\sim 0.3\,$keV. In Sec.~\ref{sec:delta}, we calculate the neutrino spectrum resulting from $p\gamma$ interactions and fit those results to the spectrum reported by the IceCube Collaboration. In Sec.~\ref{sec:igrb}, we show that the break in the diffuse neutrino spectrum at $E_{\nu}\sim 30 \, {\rm TeV}$  reduces the accompanying gamma ray flux at GeV-TeV energies, leading to consistency with measurements of the IGRB. In Sec.~\ref{sec:inSitu}, we show that if the sources are optically thick to gamma rays -- as would be expected for sources that efficiently produce neutrinos through $p\gamma$ interactions -- the contributions to the IGRB are further suppressed, especially at GeV-scale energies and above. In Sec.~\ref{sec:concl}, we summarize our results and conclusions.

\section{High-Energy Neutrinos From The Delta Resonance}\label{sec:delta}

We begin by performing a numerical calculation of the neutrino spectrum produced by high-energy protons incident on a population of target photons. To this end, we utilize the parameterization provided by \cite{Hummer:2010vx}, which is based on the results of the software package, SOPHIA~\citep{Mucke:1999yb}. To account for the effects of cosmological redshift, we integrate the parameterized spectrum over a distribution of sources, 
\begin{align}
    \frac{dN_{\nu}}{dE_{\nu}}(E_\nu) = 
    \frac{c}{4 \pi} 
    \int \frac{\dot{n}(z) \, dz}{H(z)} \bigg(\frac{dN_{\nu}}{dE'}\bigg)_{E'=E_{\nu}(1+z)},
    \label{eq:cosmology}
\end{align}
where $\dot {n}(z)$ is the comoving rate of sources as a function of redshift, $H(z) = H_0 [\Omega_M(1+z)^3+\Omega_{\Lambda}]^{1/2}$ is the Hubble parameter, and $dN_{\nu}/dE'$ is the neutrino spectrum produced per source (in units of neutrinos per energy). We adopt $H_0=67.4 \, {\rm km}/{\rm s}/{\rm Mpc}$, $\Omega_M = 0.315$, and $\Omega_{\Lambda} = 0.685$~\citep{Planck:2018vyg}. We take $\dot{n}(z)$ to be the star formation rate, as described in \cite{yuksel2008revealing}. 

\begin{figure*}[t]
    \centering
    \includegraphics[width=0.49\linewidth]{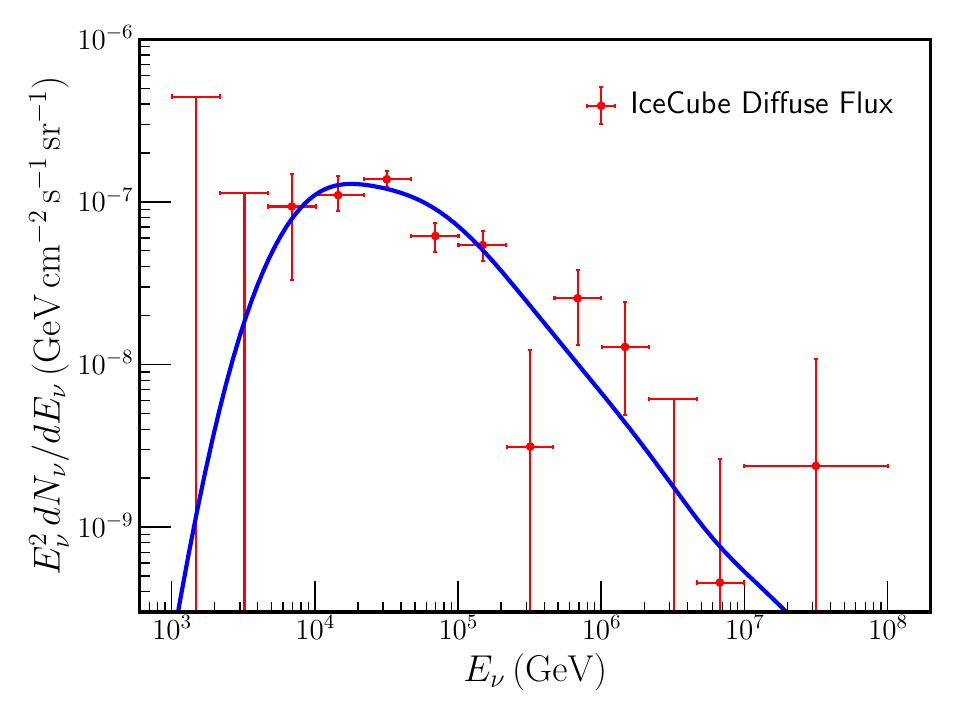}
     \includegraphics[width=0.49\linewidth]{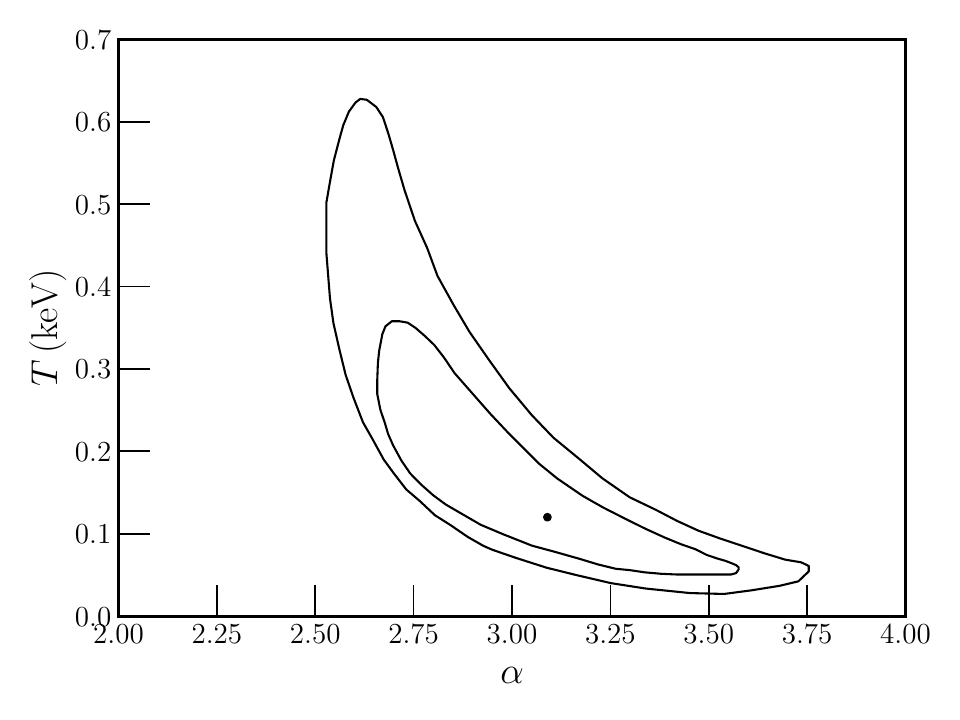}
    \caption{Left: The diffuse, all-flavor neutrino (plus antineutrino) spectrum predicted from a population of sources distributed according to the star formation rate, accelerating protons with a power-law index of $\alpha=3.1$, and incident on a blackbody spectrum of target radiation with $T = 0.12 \, {\rm keV}$. This prediction is compared to the diffuse neutrino spectrum reported by the IceCube Collaboration~\citep{abbasi2025improved}. For these parameters, the $\Delta$ resonance naturally leads to a spectral feature at $E_{\nu} \sim 30 \, {\rm TeV}$, similar to that measured by IceCube. Right: The best-fit values and 1 and 2$\sigma$ confidence contours for the proton spectral index, $\alpha$, and the temperature of the blackbody target radiation, $T$, as obtained by fitting the predicted spectrum to IceCube's measurement of the diffuse neutrino flux.}
    \label{fig:DH}
\end{figure*}

For simplicity, we adopt a proton spectrum with a power-law form, $dN_p/dE_p \propto E_p^{-\alpha}$, and a blackbody spectrum for the target radiation field. After scanning over the spectral index, $\alpha$, and the temperature of the blackbody, we find that the best fit to the diffuse neutrino spectrum reported by the IceCube Collaboration, adopting the error bars associated with their ``Combined Fit''~\citep{abbasi2025improved}, is obtained for $\alpha \approx 3.1$ and $T \approx 0.12 \, {\rm keV}$, corresponding to a mean photon energy of $\langle E_{\gamma}\rangle \approx 2.7 \, T \approx 0.3 \, {\rm keV}$. his is consistent with the findings of \citet{Winter:2026fmg}. 

The neutrino spectrum resulting from these best-fit parameters is shown in the left frame of Fig.~\ref{fig:DH}, while in the right frame we show the regions of the parameter space that are preferred by our fit at the 1 and 2$\sigma$ confidence levels. For protons spectral indices of $\alpha \sim 2.7-3.5$ and $\langle E_{\gamma} \rangle \sim 0.1-1 \, {\rm keV}$, the $\Delta$ resonance naturally leads to a spectral feature near $E_{\nu} \sim 30 \, {\rm TeV}$, consistent with that measured by IceCube. The relatively soft spectral index required by our fit is primarily determined by the shape of the neutrino flux well above the spectral break, whereas the location of the break is largely set by the characteristic energy of the target photons.

\begin{figure*}[t!]%
   \centering
\includegraphics[width=0.8\textwidth]{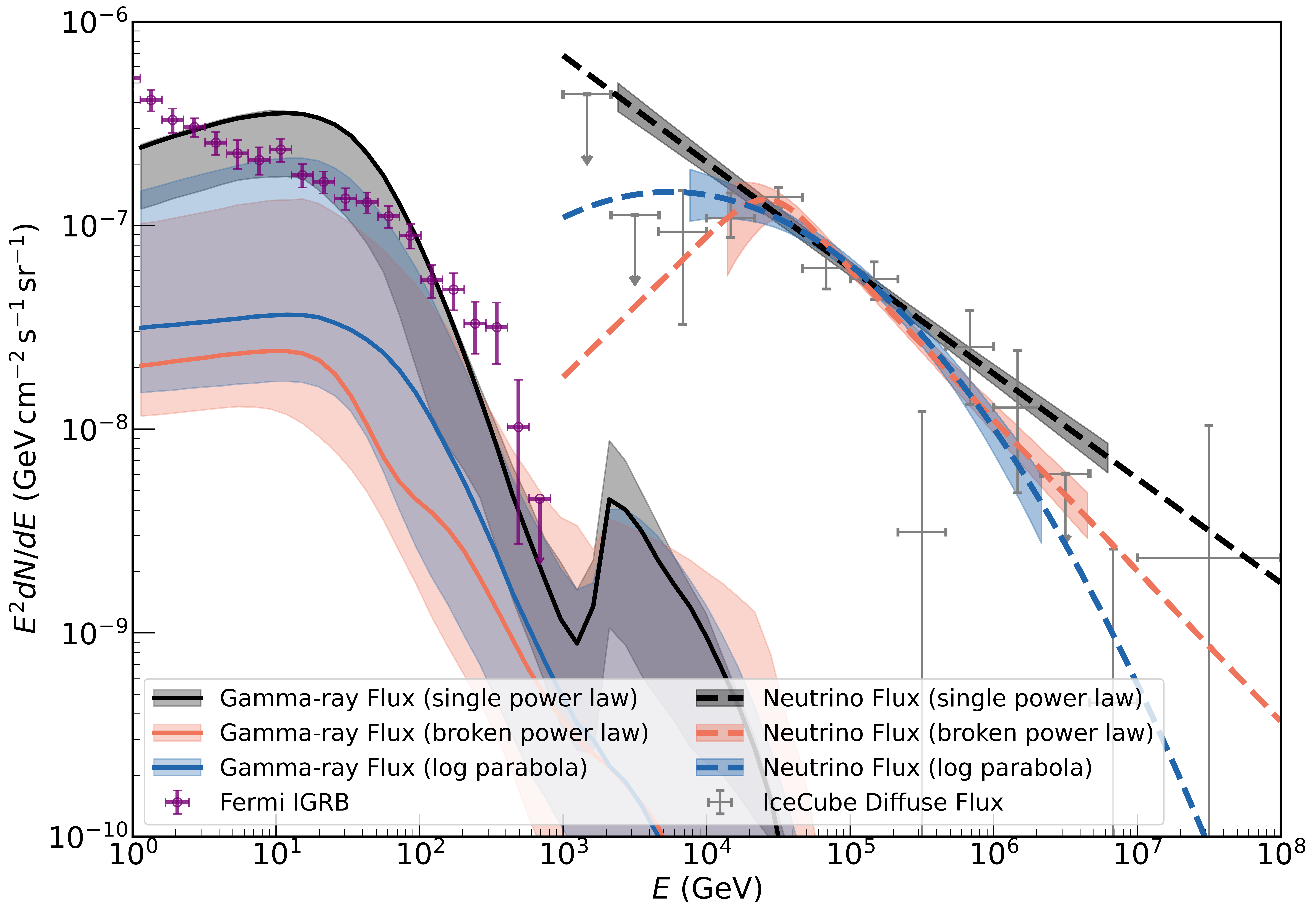}
    \caption{The spectrum of the diffuse, all-flavor neutrino (plus antineutrino) flux and that of the corresponding electromagnetic cascade, assuming that the sources of the neutrinos are transparent to gamma rays. Results are shown for single power-law (black), broken power-law (orange), and log-parabola (blue) parameterizations of the neutrino spectrum, adopting parameters as found in~\cite{abbasi2025improved}. For each parameterization, we show the predicted contribution to the isotropic gamma-ray background (IGRB) and compare this to the flux measured by {\it Fermi}-LAT~\citep{ackermann2015spectrum}. The bands surrounding these gamma-ray fluxes reflect the uncertainties in the spectral fit parameters, combined with the envelope of predictions obtained using a selection of models for the extragalactic background light. For a single (unbroken) power law extrapolated down to $E_{\nu}=1 \,{\rm TeV}$, the contribution from these sources to the IGRB  exceeds or approximately saturates the measured flux. This is in significant tension with the facts that the IGRB is known to be dominated by blazars at $E_{\gamma} > 50 \, {\rm GeV}$~\citep{Fermi-LAT:2015otn},
   and that blazars are not correlated with the neutrinos detected by IceCube~\citep{IceCube:2025tmc,Jain:2026jdh,IceCube:2016qvd,Smith:2020oac,Hooper:2018wyk}.
    In contrast, if there is a spectral break near $E_{\nu} \sim 30 \, {\rm TeV}$, as could arise from $p\gamma$ scattering through the $\Delta$ resonance, the predicted contribution to the IGRB is reduced substantially, alleviating the tension between these measurements.}
    \label{fig:exsitu}
\end{figure*}

\section{Contributions to the Isotropic Gamma-Ray Background}
\label{sec:igrb}

Neutrinos produced in astrophysical sources via charged pion production are necessarily accompanied by gamma rays from the decay of neutral pions. If sufficiently energetic, these gamma rays will undergo electron-positron pair production on the EBL, initiating an electromagnetic cascade. For  sources that are optically thin to gamma rays, the total energy flux of this cascade can be related to the diffuse neutrino spectrum according to
\begin{align}
    E_{\nu}^2
    \frac{d N_{\nu}}{d E_{\nu}}
    \simeq
    \frac{3K_{\pi}}{4}
    E_{\gamma}^2 \frac{d N_{\gamma}}{dE_{\gamma}} \Bigg|_{E_{\gamma} = 2E_{\nu}}, 
    \label{eq: nugammarelation}
\end{align}
where $K_{\pi}$ is the ratio of charged-to-neutral pions produced in the interactions, which for $p\gamma$ interactions is approximately equal to unity. The factor of $3/4$ in this expression reflects the fraction of the energy in charged pions that goes into neutrinos.  Although the injected gamma-ray spectrum is reprocessed to lower energies in optically thick environments, the total energy in these particles is conserved in a scenario where the pairs do not lose energy through other channels, allowing this relation to be used to normalize the intensity of the resulting cascade spectrum. 

To calculate the spectrum of the resulting electromagnetic cascade, we begin with the diffuse neutrino spectrum, which we parameterize either as a single power law,
\begin{align}
\frac{dN_\nu}{dE_\nu}
=
N_{\nu,0}
\, E_\nu^{-\gamma}, 
\label{eqPL}
\end{align}
as a broken power law,
\begin{align}
\frac{dN_\nu}{dE_\nu}
=
N_{\nu,0}
\begin{cases}
\left(\dfrac{E_\nu}{E_{\rm break}}\right)^{-\gamma_1}, & E_\nu < E_{\rm break}, \\[1em]
\left(\dfrac{E_\nu}{E_{\rm break}}\right)^{-\gamma_2}, & E_\nu > E_{\rm break},
\end{cases}
\label{eqBPL}
\end{align}
or as a log parabola,
\begin{align}
   \frac{dN_{\nu}}{dE_{\nu}}
   =
   N_{\nu, 0}
   \bigg(
   \frac{
   E_{\nu}
   }{
   E_{\nu, 0}
   }
   \bigg)
   ^{-\alpha-\beta \log(E_{\nu}/E_{\nu, 0})}
   .
   \label{eqNuspectrum}
\end{align}
In each case, we adopt values for these parameters ($N_{\nu,0}, \gamma, \gamma_1, \gamma_2, E_{\rm break}, E_{\nu,0}, \alpha, \beta$) as reported by the IceCube Collaboration, from their ``Combined Fit'' analysis~\citep{abbasi2025improved}. These values are tabulated in Table~\ref{tab:combined_fit_parameters}. Note that energy ranges of broken power-law and log-parabola models stated refer to the analysis energy limits, and it will be used to plot the central values in Fig.~\ref{fig:exsitu}. On the other hand, the upper limits are computed by setting the minimum energies to 1 TeV.

\begin{table}[htbp]
\centering
\caption{Best-fit spectral parameters for the ``Combined Fit'' single power-law (SPL), broken power-law (BPL) and log-parabola injection models. The energy range of each model is also listed. Note that the neutrino flux normalization factors are expressed in units of $C_0 \equiv 3  \times 10^{-18}\,\mathrm{GeV^{-1}\,cm^{-2}\,s^{-1}\,sr^{-1}}$.}
\label{tab:combined_fit_parameters}

\setlength{\tabcolsep}{5pt}
\begin{tabular}{llc}
\hline
\hline
Model & Parameter & Value \\
\hline
\hline
\multirow{3}{*}{SPL}
& $N_{\nu,0}/C_0$
& $1.80^{+0.13}_{-0.16} $ \\ 
& $\gamma$ 
& $2.52^{+0.036}_{-0.038}$ \\
& Energy range
& $1~\mathrm{TeV}-6.4~\mathrm{PeV}$ \\
\hline

\multirow{5}{*}{BPL}
& $N_{\nu,0}/C_0$
& $1.77^{+0.19}_{-0.18}  $ \\ 
& $\gamma_1$
& $1.31^{+0.51}_{-1.30}$ \\
& $\gamma_2$
& $2.735^{+0.067}_{-0.075}$ \\
& $\log_{10}(E_{\mathrm{break}}/\mathrm{GeV})$
& $4.39 \pm 0.1$ \\
& Energy range
& $13.7~\mathrm{TeV}-4.7~\mathrm{PeV}$ \\
\hline

\multirow{5}{*}{log parabola}
& $N_{\nu,0}/C_0$
& $2.13^{+0.16}_{-0.19}$ \\ 
& $\alpha$
& $2.572^{+0.062}_{-0.053}$ \\
& $\beta$
& $0.228^{+0.098}_{-0.067}$ \\
& $E_{\nu,0}$
& $100~\mathrm{TeV}$  \\
& Energy range
& $7.5~\mathrm{TeV}-2.2~\mathrm{PeV}$ \\
\hline
\end{tabular}
\end{table}


From the injected spectrum of gamma rays, we calculate the contribution to the IGRB using the publicly available code, CRPropa 3.2~\citep{CRPropa:2022ovg}, which includes the processes of electron-positron pair production and inverse Compton scattering. For the EBL, we  adopt the model of~\cite{ dominguez2011extragalactic} as our default choice, but scan over a selection of seven different models~\citep{kneiske2004implications, stecker2006intergalactic, franceschini2008extragalactic,gilmore2012semi, saldana2021observational, finke2022modeling}. We construct a final uncertainty band that reflects these variations as well as those associated with the parameters of IceCube's spectral fit. 

For the case of a single (non-broken) power law, extrapolated down to an energy of $E_{\nu} = 1 \, {\rm TeV}$, the predicted gamma-ray flux contributes $\sim 24-100$\% to the measured IGRB~\citep{ackermann2015spectrum} in the energy range of $E_{\gamma} \sim85-820$ GeV, while exceeding this gamma-ray background between energies of $E_{\gamma} \sim3-60$ GeV 
(see Fig.~\ref{fig:exsitu}). Note that there is a feature in the gamma-ray spectrum at $E_{\gamma} \sim 10-50 \, {\rm TeV}$ which is due to sources that lie within $\sim 50 \, {\rm Mpc}$ and which do not experience significant gamma-ray attenuation. 

\begin{figure*}%
    \centering
\includegraphics[width=0.8\textwidth]{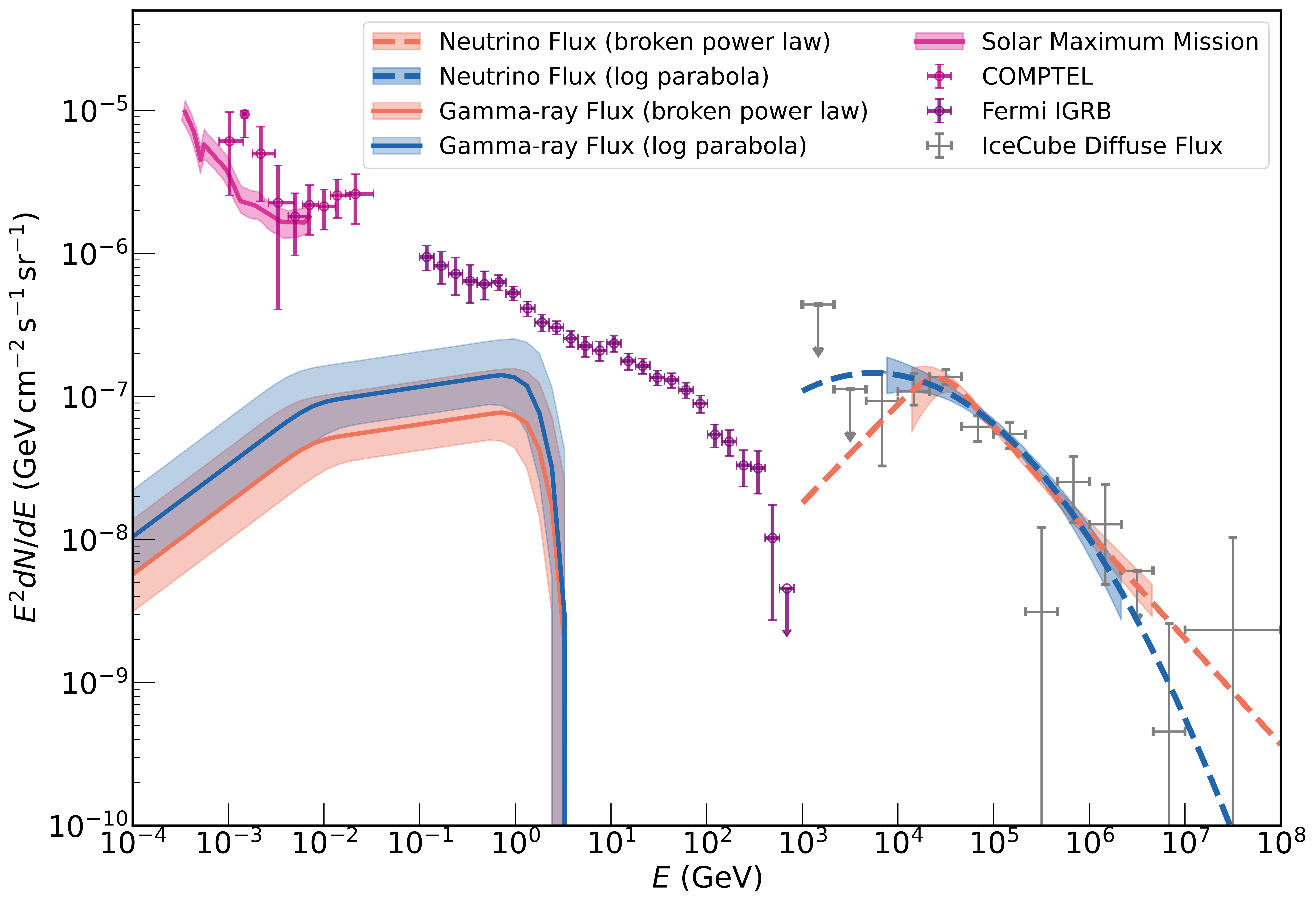}
    \caption{The spectrum of the diffuse, all-flavor neutrino (plus antineutrino) flux and that of the corresponding electromagnetic cascade, assuming that the sources of the neutrinos are opaque to gamma rays (see text for details). Results are shown for broken power-law (orange) and log-parabola (blue) parameterizations of the neutrino spectrum, adopting parameters as found in~\cite{ackermann2015spectrum}. For each parameterization, we show the predicted contribution to the isotropic gamma-ray background (IGRB) and compare this to the flux measured by {\it Fermi}-LAT~\citep{ackermann2015spectrum}, COMPTEL~\citep{weidenspointner2001comptel}, and the Solar Maximum Mission (SMM)~\citep{watanabe2000mev}. The bands around these gamma-ray fluxes reflect the uncertainties in the spectral-fit parameters, combined with an envelope of predictions obtained for a selection of extragalactic background light models and target photon spectra. These predicted contributions to the gamma-ray background are all well below the measured intensity of the IGRB.}
    \label{fig:insitu}
\end{figure*}

Blazars are known to provide the dominant contribution to the IGRB at energies above $E_{\gamma} \sim 50 \, {\rm GeV}$~\citep{Fermi-LAT:2015otn}, along with significant contributions from non-blazar AGN and starforming galaxies~\citep{di2014diffuse, roth2021diffuse,Ajello:2020zna,Fornasa:2015qua,Ajello:2015mfa,Hooper:2016gjy,DiMauro:2015tfa}. Furthermore, the directions of IceCube's neutrinos do not correlate with those of known gamma-ray blazars~\citep{IceCube:2025tmc,Jain:2026jdh,IceCube:2016qvd,Smith:2020oac,Hooper:2018wyk}. Taken together, these considerations indicate that most of the IGRB cannot originate from the sources of IceCube's diffuse neutrino flux. From this, and as argued in \cite{Murase:2015xka},~\cite{ Hooper:2016jls}, and~\cite{fang2022tev}, we conclude that the diffuse neutrino spectrum must either originate from sources that are optically thick to gamma rays (see Sec.~\ref{sec:inSitu}), or that the diffuse neutrino flux must feature a spectral break at an energy no lower than $E_{\nu} \sim 2-4 \, {\rm TeV}$.

If the pion spectrum -- and therefore the neutrino and gamma-ray spectra -- deviates from a single power law below TeV-scale energies, this tension with the measured intensity of the IGRB can be straightforwardly alleviated. For the spectrum shown in Fig.~\ref{fig:DH}, for example, the electromagnetic cascade contributes only $\sim 10\%$ of the measured IGRB at 200 GeV. This conclusion does not necessarily rely on the physics of the $\Delta$ resonance, or even on $p\gamma$ interactions. More generally, the tension between the diffuse neutrino flux and the IGRB can be reconciled in a broad range of scenarios in which the diffuse neutrino spectrum does not extend with a single power-law down to energies below the TeV-scale. Such a spectral shape could also arise, for example, from a break in the injected proton spectrum associated with magnetic reconnection \citep{Fiorillo:2023dts,Fang:2024wmf}, or, in environments with extreme magnetic fields, from the suppression of secondary particles (muons, pions, and kaons) due to synchrotron losses~\citep{Winter:2026fmg,Blanco:2025zqo}.

In Fig.~\ref{fig:exsitu}, we show that if we fit IceCube's diffuse neutrino spectrum with either a broken power-law or a log-parabola parameterization, the predicted contributions to the IGRB that are more consistent with, or lie below, the flux measured by {\it Fermi}-LAT. In particular, for a broken power-law injection model, we find that the sources of the diffuse neutrino flux produce $\sim 5-73\%$ of the total IGRB between $1-100$ GeV. This contribution is comparable to the cascade gamma-ray flux predicted in the minimal $p\gamma$ scenario of \citet{Murase:2015xka}, assuming that the neutrino spectrum extends down to $\sim1$~TeV. In the log-parabola case, the fraction is much larger, saturating the IGRB and producing more than $\sim 100\%$ in the 10-100 GeV energy range. Above 100 GeV, the predicted cascade flux comprises $\sim 6-72\%$ and $\sim 8-100\%$ of the IGRB data for these two models, respectively.


\section{Gamma Rays From Optically Thick Sources}
\label{sec:inSitu}

In the previous section, we assumed that the sources of IceCube's diffuse neutrino flux are transparent to gamma rays. If these sources efficiently produce neutrinos through $p\gamma$ interactions, however, they must contain dense fields of radiation, causing high-energy gamma rays to undergo electron-positron pair production {\it in situ}. This process is expected to reprocess any electromagnetic emission to lower energies, where the source becomes transparent to such photons.

To calculate the spectral shape of the reprocessed electromagnetic emission, we adopt a radiation field with a dichromatic spectrum, $\epsilon\,dn/d\epsilon$, peaking at $\epsilon_l$ and $\epsilon_h$. This choice is motivated by the fact that many astrophysical sources contain well-separated ``hot'' and ``cool'' radiation components. We take $\epsilon_h$ to lie within the range $0.2-0.5\, {\rm keV}$ in order to produce a spectral break at $E_{\nu} \sim 30 \, {\rm TeV}$ (as shown in Fig.~\ref{fig:DH}). This choice is further motivated by AGN environments such as NGC 1068 and TXS 0506+056, where soft X-ray photons from the corona provide sufficient photohadronic opacity to facilitate the efficient production of neutrinos~\citep{khatee2025ngc}. We set $\epsilon_l = 1 \,{\rm eV}$ as a representative energy for the thermal emission from an AGN's accretion disk, although the precise value of this quantity has little impact on the resulting cascade spectrum.

For a source that is optically thick to gamma rays, the electromagnetic cascade will develop fully, resulting in a spectral shape of the following form~\citep{berezinsky1975cosmic, berezinsky2016high, fang2022tev}:
\begin{equation}\label{eqn:analySpectrum}
\frac{dN_{\rm cas}}{dE_\gamma}=\begin{cases}
			K  (E_\gamma/{\cal E}_X)^{-3/2}, &    E_\gamma < {\cal E}_X \\
             K  (E_\gamma/{\cal E}_X)^{-2}, &  {\cal E}_X < E_\gamma < {\cal E}_\gamma \\
            0, & E_\gamma > {\cal E}_\gamma,
		 \end{cases}
\end{equation}
where ${\cal E}_\gamma \equiv 4 m_e^2 / \epsilon_h$ corresponds to the threshold gamma-ray energy for pair production and ${\cal E}_X \equiv 4 ({\cal E}_\gamma / 2 m_e)^2 \epsilon_l/3$ is the energy of the photons up-scattered by the final generation of electron-positron pairs.

We show the results of this calculation in Fig.~\ref{fig:insitu}, parameterizing the diffuse neutrino spectrum as either a broken power law or a log parabola~\citep{abbasi2025improved}. For each case, we show the spectrum of the electromagnetic cascade predicted from these sources, as compared to the {\it Fermi} measurement of the IGRB~\citep{ackermann2015spectrum}, as well as measurements of the gamma-ray background by COMPTEL~\citep{weidenspointner2001comptel}, and the Solar Maximum Mission (SMM)~\citep{watanabe2000mev}. The uncertainty bands on the cascade flux in this figure reflect the variation across EBL models, spectral-fit parameters, and $\epsilon_h=0.2-0.5 \, {\rm keV}$.

The contribution to the IGRB from this population of optically thick sources is well below the measured flux. For example, in the $E_{\gamma} \sim 0.1 - 1$ GeV range, the predicted contribution to the gamma-ray background constitutes only $\sim 12-28 \%$ ($\sim 7-16 \%$) of the total IGRB for a log-parabola (broken power-law) injection model.

\section{Summary and Conclusions}
\label{sec:concl}

Recent measurements of the diffuse high-energy neutrino flux by the IceCube Collaboration have revealed evidence for a spectral feature at $E_{\nu} \sim 30 \, {\rm TeV}$. In this letter, we have explored the origin and implications of this spectral break in the context of photohadronic ($p\gamma$) interactions. In particular, we have shown that the $\Delta$-baryon resonance provides a natural mechanism for generating a break at the observed energy. Interactions between protons with energies of $E_p \sim 0.1$–$1 \, {\rm PeV}$ and target photons in the soft X-ray band, $E_{\gamma} \sim 0.1$–$1 \,{\rm keV}$, yield neutrinos with energies near $E_{\nu} \sim 30 \, {\rm TeV}$, in good agreement with the IceCube data. A simple model consisting of a power-law proton spectrum interacting with a thermal radiation field provides an excellent fit to the observed diffuse neutrino spectrum, with best-fit parameters of $\alpha \approx 3.1$ and $T \approx 0.1 \,{\rm keV}$.

We have further examined the implications of this spectral feature for contributions from neutrino sources to the isotropic gamma-ray background (IGRB). In scenarios in which the neutrino spectrum extends as an unbroken power law to low energies, the associated electromagnetic cascades are expected to exceed the observed IGRB, leading to a well-known tension. We have demonstrated that the presence of a spectral break at $E_{\nu} \sim 30 \, {\rm TeV}$  reduces the predicted gamma-ray flux, bringing it into consistency with the measurements of {\it Fermi}-LAT. 

We have also considered the case in which the sources of IceCube's neutrinos are optically thick to gamma rays, as would be expected for environments with high photohadronic efficiency. In this scenario, the electromagnetic emission is reprocessed {\it in situ}, leading to a cascade spectrum that peaks at MeV–GeV energies. We find that the resulting contribution to the IGRB is subdominant across a wide range of model parameters, further relieving tensions with {\it Fermi}-LAT.

Future measurements by IceCube and next-generation neutrino observatories will further clarify the shape of the diffuse neutrino spectrum, including the presence of any spectral features. Combined with improved gamma-ray observations, these data will provide increasingly stringent constraints on the origin of high-energy neutrinos and the environments in which they are produced. In fact, there are about a thousand active galaxies with an X-ray flux larger than NGC 1068~\citep{Urry:1995mg}. If these were to accommodate the diffuse flux of extragalactic cosmic rays, they could collectively preserve the $\Delta$ resonance feature in the diffuse flux.

\vspace{1cm}

This work has been supported by the Office of the Vice Chancellor for Research at the University of Wisconsin--Madison with funding from the Wisconsin Alumni Research Foundation.  F.H. and A.K.Z. are also supported by the National Science Foundation under grants~PHY-2209445 and OPP-2042807, and by the Balzan Foundation. K.F. acknowledges support from the National Science Foundation (PHY-2238916, PHY-2514194), a Sloan Research Fellowship, and the Simons Foundation (00001470). 


\bibliography{biblio}{}
\bibliographystyle{aasjournal}

\end{document}